# Fixed Volume Effect on Polar Properties and Phase Diagrams of Ferroelectric Semi-ellipsoidal Nanoparticles

**Victoria V. Khist[1], Anna N. Morozovska[2*], Maxim V. Silibin[3], Yevhen M. Fomichov[4], George S. Svechnikov[5], Andrei L. Kholkin[6], Vladimir V. Shvartsman[7], Dmitry V. Karpinsky[3,8] and Eugene A. Eliseev[4]**

[1]Institute of Magnetism, National Academy of Sciences of Ukraine and Ministry of Education and Science of Ukraine, Prospekt Vernadskogo 36a, 03142 Kyiv, Ukraine,

[2]Institute of Physics, National Academy of Sciences of Ukraine, 46, pr. Nauky, 03028 Kyiv, Ukraine

[3] National Research University of Electronic Technology "MIET", Moscow, Zelenograd, Russia,

[4] Institute for Problems of Materials Science, National Academy of Sciences of Ukraine, 3, Krjijanovskogo, 03142 Kyiv, Ukraine,

[5]Sikorsky Kyiv Polytechnic Institute, Prospekt Pobedi 37, Kyiv 03052, Ukraine

[6] Department of Physics and CICECO – Materials Institute of Aveiro, University of Aveiro, 3810-193 Aveiro, Portugal

[7] Institute for Materials Science and Center for Nanointegration Duisburg-Essen (CENIDE), University of Duisburg-Essen, Universitatsstrasse 15, 45141 Essen, Germany

[8] Scientific-Practical Materials Research Centre of NAS of Belarus, Minsk, Belarus

For advanced applications in modern industry it is very important to reduce the volume of ferroelectric nanoparticles without serious deterioration of their polar properties. In many practically important cases fixed volume (rather than fixed size) corresponds to realistic technological conditions of nanoparticles fabrication. The letter is focused on the theoretical study of the behavior of ferroelectric polarization, paramagnetoelectric coefficient and phase diagrams of semi-ellipsoidal nanoparticles with fixed volume $V$. Our approach combines the Landau-Ginzburg-Devonshire phenomenology, classical electrostatics and elasticity theory. Our results show that the size effects of the phase diagrams and polarization of semi-ellipsoidal $BiFeO_3$ nanoparticles nontrivially depends on $V$. These findings provide a path to optimize the polar properties of nanoparticles by controlling their phase diagrams at a fixed volume.

---

[*] corresponding author, e-mail: anna.n.morozovska@gmail.com

# I. INTRODUCTION

Overall miniaturization of functional ferroic materials is highly attractive for modern industry, however, the estimation of the physical properties at nanoscale level is a difficult scientific task. The resulting alteration of the physical properties with a reduction of the sample size down to a nanoscale range is currently an area of intensive study [1, 2, 3, 4]. Modern fabrication technologies for microactuators, microwave phase shifters, infrared sensors, transistor applications, energy harvesting devices etc. need comprehensive understanding about the correlation between samples sizes and their geometry, on the one hand, and polar and magnetic order, domain sizes, domain wall thickness and other parameters, on the other hand [1-4].

Reduction of the ferroic dimension down to a nanoscale level leads to drastic changes (including notable enhancement or suppression) of its polar, magnetic, and magnetoelectric properties. Multiple examples have been found experimentally and explained theoretically, e.g. in Rochelle salt [5, 6, 7, 8, 9], $BaTiO_3$ [10, 11, 12, 13, 14, 15], $Pb(Zr_{1-x}Ti_x)O_3$ [16, 17, 18], $KTa_{1-x}Nb_xO_3$ [19, 20, 21, 22], $SrTiO_3$ [23], and $SrBi_2Ta_2O_9$ [24, 25, 26] ferroelectric nanoparticles, as well as for pristine [27, 28, 29, 30, 31] and rare-earth doped [32, 33, 34, 35, 36, 37] $BiFeO_3$ multiferroic nanoparticles, nanograins, nanoislands [38] and their self-assembled arrays [39, 40, 41].

$BiFeO_3$ based materials continue to attract significant scientific interest due to the number of phase transitions and related changes in the unique multiferroic properties [42, 43, 44, 45, 46, 47, 48, 49] under different stimuli such as chemical doping, pressure, temperature and radiation. However, most studies [27, 29-41] of the size-dependent properties of nanosized ferroelectric antiferromagnet $BiFeO_3$ unequivocally show the decrease of polar and antifferomagnetic transition temperatures with a reduction of grain size; e.g. a Neel transition temperature is around 550 K for (10 – 5) nm $BiFeO_3$ grains [27] (in comparison with 650 K for a bulk sample). Pronounced effects on remanent polarization, dielectric permittivity, coercive field and ferroelectric phase transition temperature have also been observed for $BiFeO_3$ nanoparticles with effective grain sizes below 50 nm [50, 51, 52]. All these results suggest that at the nanoscale level the properties of $BiFeO_3$ are exceptional and complex. In spite of a number of studies devoted to size effects on physical properties of $BiFeO_3$-based materials [27, 29-41] there is no self-consistent model describing the evolution of the polarization as a function of grain volume in the nanoscale range for different grain geometries.

Recently [29] the size effects on phase diagrams, ferroelectric, and magnetoelectric properties of semi-ellipsoidal $BiFeO_3$ nanoparticles clamped to a rigid conductive substrate have been studied using the Landau-Ginzburg-Devonshire (LGD) phenomenology [53, 54].

Nanoparticles of semi-ellipsoidal shape are considered as model objects to study size effects on physical properties of ferroic nano-islands. BiFeO$_3$ nano-islands and their self-assembled arrays can be formed on anisotropic substrates by different low-damage fabrication methods [39, 40, 41].

The above state-of-art analyses motivated us to conduct a theoretical study of the size effects on ferroelectric properties of semi-ellipsoidal BiFeO$_3$ nanoparticles under a fixed volume condition [**Fig. 1**] corresponding to the realistic experiments [39, 40, 41]. We have used LGD phenomenology combined with the classical electrostatics and elasticity theory.

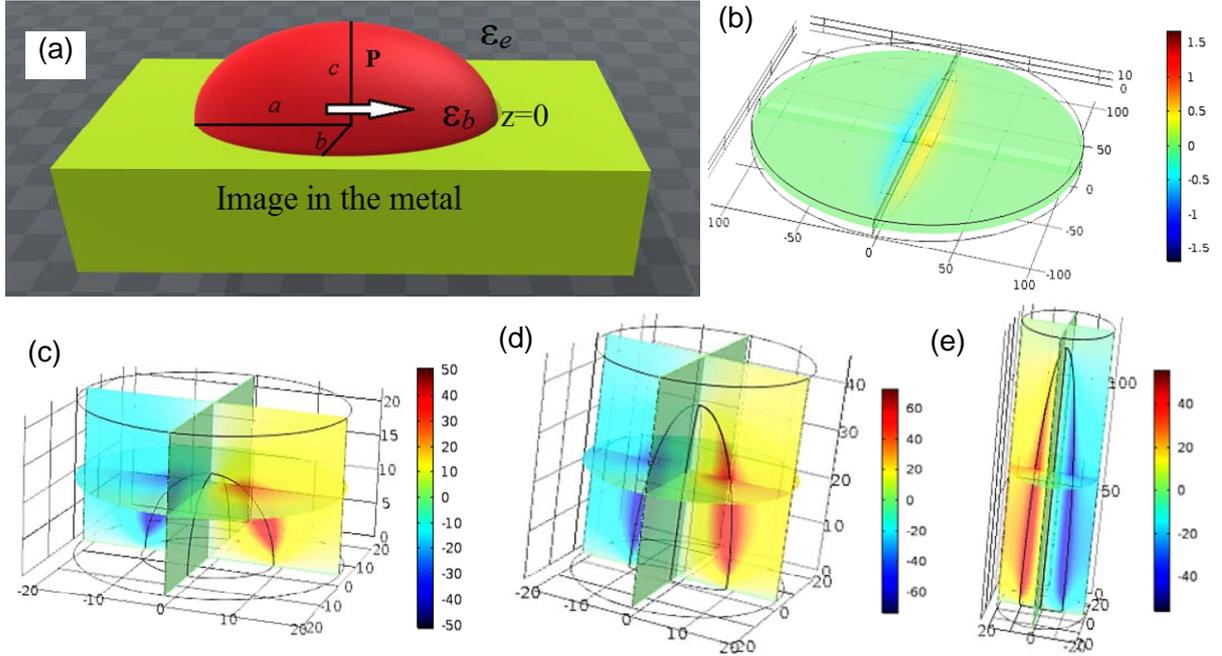

**FIG. 1.** (a) Semi-ellipsoidal uniformly polarized ferroelectric nanoparticles are clamped to a rigid conducting substrate electrode. The one-component ferroelectric polarization **P**(**r**) is directed along the X-axes. Semi-ellipsoid height is denoted by $c$ and lateral semi-axes by $a$ and $b$, respectively. Parts **(b)-(e)** show the electric potential distribution inside the particles, which have different aspect ratios $a/b$ =0.1, 1, 3, 10 and the same volume. Color scale is in Volts.

## II. PROBLEM STATEMENT AND BASIC EQUATIONS

We consider ferroelectric nanoparticles in the form of semi-elliptical islands precipitated on a rigid conducting substrate electrode. The ellipsoid has different values of semi-axis length, $a$, $b$ and $c$ along the X-, Y- and Z-axis, respectively [see **Fig. 1(a)**]. The sizes can vary, while the particle volume $V = (2/3)\pi abc$ is fixed to satisfy technological conditions of the nanoparticle preparation by e.g. laser ablation. We denote $\varepsilon_b$ and $\varepsilon_e$ as the isotropic dielectric permittivity of ferroelectric background [55] and external media, respectively. The dependences of the $x$- and $y$-components of the electric polarizations on the electric field **E** are linear, $P_{y,z} = \varepsilon_0(\varepsilon_b - 1)E_{y,z}$, and the polarization $x$-component contains the ferroelectric ($P$) and background ($\sim E_x$) contributions, $P_x = P + \varepsilon_0(\varepsilon_b - 1)E_x$, where $\varepsilon_0$ is the universal dielectric constant. The ferroelectric contribution, $P(\mathbf{r}, E_3)$, was determined from the LGD equation inside a nanoparticle [29],

$$\alpha_P P + \beta_P P^3 + \gamma_P P^5 - g_{33mn} \frac{\partial^2 P}{\partial x_m \partial x_n} = E_x, \qquad (1)$$

where the coefficient $\alpha_P(T) = \alpha_P^{(T)}(T - T_C)$, $T$ is the absolute temperature, and $T_C$ is the bulk Curie temperature of the paraelectric-to-ferroelectric phase transition. The parameters $\beta_P$ and $\gamma_P$ are the coefficients of LGD potential expansion on the polarization powers. Boundary conditions for the polarization $P_3$ at the particle surface S are regarded to be natural, $(\partial P_x / \partial \mathbf{n})|_S = 0$.

Electric field $E_i$ is defined via electric potential as $E_i = -\partial \varphi / \partial x_i$. For a ferroelectric particle without free charges, the electric potential $\varphi$ can be found self-consistently from the Laplace equation outside the nanoparticle ($\varepsilon_0 \varepsilon_e \Delta \varphi = 0$) and Poisson equation inside the nanoparticle ($\varepsilon_0 \varepsilon_b \Delta \varphi = \partial P / \partial x$). The corresponding electric boundary conditions are the continuity of potential at the particle surface $S$, $(\varphi_e - \varphi_i)|_S = 0$, and the difference of normal components of electric displacements, which is equal to the surface screening charge density at the particle surface S, $(\mathbf{D}_e - \mathbf{D}_i)\mathbf{n} + \varepsilon_0 \frac{\varphi_i}{\lambda}\Big|_S = 0$, where $\mathbf{D}_i = \varepsilon_0 \varepsilon_b \mathbf{E}_i + \mathbf{P}$ inside the particle and $\mathbf{D}_e = \varepsilon_0 \varepsilon_e \mathbf{E}_e$ outside the particle; $\lambda$ is the surface screening length. The subscript "$i$" corresponds to the electric field or potential inside the particle and "$e$" – outside the particle. The potential is constant at the particle-electrode interface, i.e. $\varphi_i|_{z=0} = 0$.

We calculated the spatial distribution and the average electric field inside the BiFeO$_3$ particles using finite element modeling (FEM), the material parameters of which are listed in **Table I**. The parameters of BiFeO$_3$ are collected from Refs.[42-49, 29].

Table I. Parameters of bulk BiFeO$_3$ used in our calculations

| Parameter | SI units | Value for BiFeO$_3$ |
|---|---|---|
| Spontaneous polarization $P_S$ | C/m$^2$ | 1 |
| Electrostriction coefficient $Q_{12}$ | m$^4$/C$^2$ | −0.016 |
| Electrostriction coefficient $Q_{11}$ | m$^4$/C$^2$ | +0.032 |
| Background permittivity $\varepsilon_b$ | dimensionless | 10 |
| Ambient permittivity $\varepsilon_e$ | dimensionless | 1 |
| Gradient coefficient $g_{11}$ | m$^3$/F | $10^{-10}$ |
| LGD coefficient $\alpha_S$ | m$^2$/F | $10^{-4}$ |
| LGD coefficient $\beta$ | J m$^5$/C$^4$ | $10^7$ |
| LGD coefficient $\alpha$ | m/F | $-10^7$ (at 300 K) |

| Ferroelectric Curie temperature $T_c$ | K | 1100 |
|---|---|---|
| Temperature coefficient $\alpha_T$ | m/(K F) | $0.9 \times 10^6$ |
| Antiferromagnetic Neel temperature | K | 650 |
| Surface screening length $\lambda$ | nm | $10^{-3}$ to $10^2$ |
| Universal dielectric constant $\varepsilon_0$ | F/m | $8.85 \times 10^{-12}$ |

By analyzing the FEM results in Ref. [29] we derived sufficiently accurate analytical expression for the transition temperature from the ferroelectric (FE) to the paraelectric (PE) phase $T_{cr}(a,b,c)$:

$$T_{cr}(a,b,c) = T_C - \frac{n_d(a,b,c)}{\alpha_T \varepsilon_0}. \qquad (2)$$

The effective depolarization factor $n_d(a,b,c)$ in Eq.(2) depends on the semi-ellipsoid geometry as listed in Ref. [29]. The average spontaneous polarization that is nonzero in the temperature range $T < T_{cr}(a,b,c)$,

$$P_S = \sqrt{\frac{\alpha_T}{\beta}(T_{cr}(a,b,c) - T)}. \qquad (3)$$

In expressions (2)-(3) the sizes *a*, *b* and *c* are related by the fixed volume condition, $V = (2/3)\pi abc$, i.e. only 2 of them can be varied independently. Rewriting the expressions for $n_d(a,b,c)$ from Ref. [29] vs. the volume $V = (2/3)\pi abc$ and ratio $a/b = \gamma$, we get that $c = \frac{3V\gamma}{2\pi a^2}$ and so

$$n_d(a,V,\gamma) = \frac{\lambda n_\infty(a,V,\gamma)}{\lambda + R(a,V,\gamma) n_\infty(a,V,\gamma)}. \qquad (4)$$

$$R \approx a\left(0.62 + 0.19\gamma + 0.52\frac{a^3}{V}\frac{1}{\gamma}\right), \qquad (5)$$

$$n_\infty \approx \frac{1}{(\varepsilon_b + \varepsilon_e \gamma)}\left(1 + 0.4\frac{a^3}{V}\frac{1}{\gamma} + \left(\frac{2\pi a^3}{3V}\right)^2 \frac{1}{\gamma^3 + 0.075\gamma^4}\right)^{-1}. \qquad (6)$$

### III. SIZE EFFECTS OF POLARIZATION AND PHASE DIAGRAMS AT FIXED VOLUME

**Figure 2** shows the phase diagram of semi-ellipsoidal BiFeO$_3$ nanoparticles. *Y*-axis is the relative temperature $T/T_C$ and X-axis is the nanoparticle volume *V*. Different curves are calculated for several values of the aspect ratio *a/b* =0.1, 1 3, 10. The boundary between paraelectric (**PE**) and ferroelectric (**FE**) phases, given by the critical temperature of the size-induced phase transition $T_{cr}(a,b,c)$, monotonously increases with the increasing *a/b* ratio. The size effect manifests itself in the ferroelectricity disappearance at a critical volume $V_{cr}(a,b)$ for which $T_{cr} = 0$, and in the monotonous increase of the transition temperature with the volume increase, followed by its further saturation to $T_C$ for the $V > 10^8$ nm$^3$.

The series of curves in **Fig. 2** show the influence of the particle volume $V$ and the aspect ratio $a/b$ (proportional to the size $a$ in the direction of spontaneous polarization) on the ferroelectric phase transition. The transition temperature monotonically increases and the critical volume $V_{cr}$ decreases with increasing $a/b$ ratio (see the sequence of black, red, blue, and magenta curves). At the minimal $a/b$ ratio the phase transition occurs at the largest value of $V$ (see the black curve). This is due to the influence of the depolarization field directed along the $x$-axis that is maximal for the smallest $a$-size. The higher is the aspect ratio $a/b$, the smaller is the depolarization field, and hence the higher the transition temperature and the smaller the critical size. For the maximal value of $a/b$, the phase transition occurs at the smallest value of $V$ (see the magenta curve). It should also be noted that the critical volume for this curve is less than 1 nm$^3$, which lies beyond the limits of continuous theory applicability (every size should be at least one order of magnitude greater that the lattice constant ~ 0.4 nm).

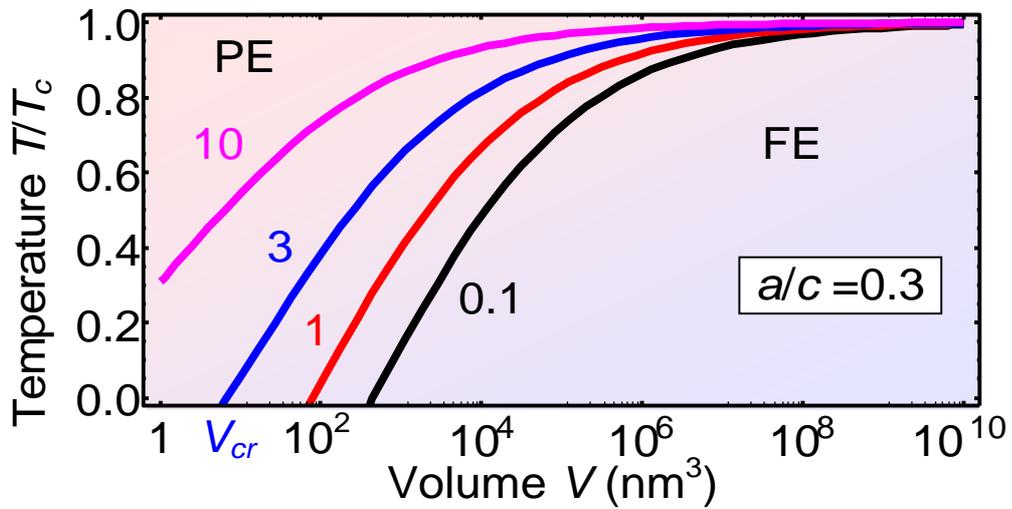

**FIG. 2.** Phase diagrams of BiFeO$_3$ nanoparticle in coordinates "temperature $T$ – volume of semi-ellipsoidal particle $V$" calculated for fixed aspect ratio $a/c$ =0.3 and different aspect ratios $a/b$ =0.1, 1, 3, 10 (see numbers near the curves) and screening length $\lambda$=1 nm.

Phase diagrams of semi-ellipsoidal BiFeO$_3$ nanoparticles in the coordinates: "relative temperature $T/T_C$" and "length of the particle semi-axis $a$", are shown in **Figs.3(a)-(c)** for the fixed particle volume $V = 5\times10^3$, $5\times10^4$ and $5\times10^6$ nm$^3$, respectively. Different curves on each panel are calculated for several values of the aspect ratio $a/b$ =0.1, 1, 3, 10. The ferroelectricity disappearance at the critical size $a_{cr}(b,c)$, for which $T_{cr} = 0$ is followed by the monotonic increase of the transition temperature with the size $a$ increase and its further saturation to $T_C$ for the sizes $a > (50-100)$ nm.

We found that the boundary between the PE and FE phases nontrivially depends on the aspect ratio $a/b$. Comparison of the curves in plots **3(a) – 3(c)** calculated for the fixed $V$ shows that the phase diagram is the most nontrivial for $V=5\times10^3$ nm$^3$, where there are multiple intersections of the curves calculated for different aspect ratios $a/b$. One intersection of the curve for $a/b=10$ (magenta curve) and $a/b=0.1$ (black curve) corresponds to $a \approx 5$ nm. Another intersection of the curves calculated for $a/b=0.1$ (black curve) and $a/b=3$ (blue curve) takes place at $a \approx 5$ nm. The curve for $a/b=1$ (red) intersects the curve for $a/b=3$ (blue) at $a \approx 10$ nm, and the curve for $a/b=10$ at $a \approx 13$ nm. The shape of the curves for $V=5\times10^4$ nm$^3$ is qualitatively the same as for $V=5\times10^3$ nm$^3$, but differs quantitatively, because the smallest semi-axes length, when an intersection occurs between the transition line for $a/b=0.1$ (black curve) with the curves for other particle aspect ratio, is $a \approx 8$ nm.

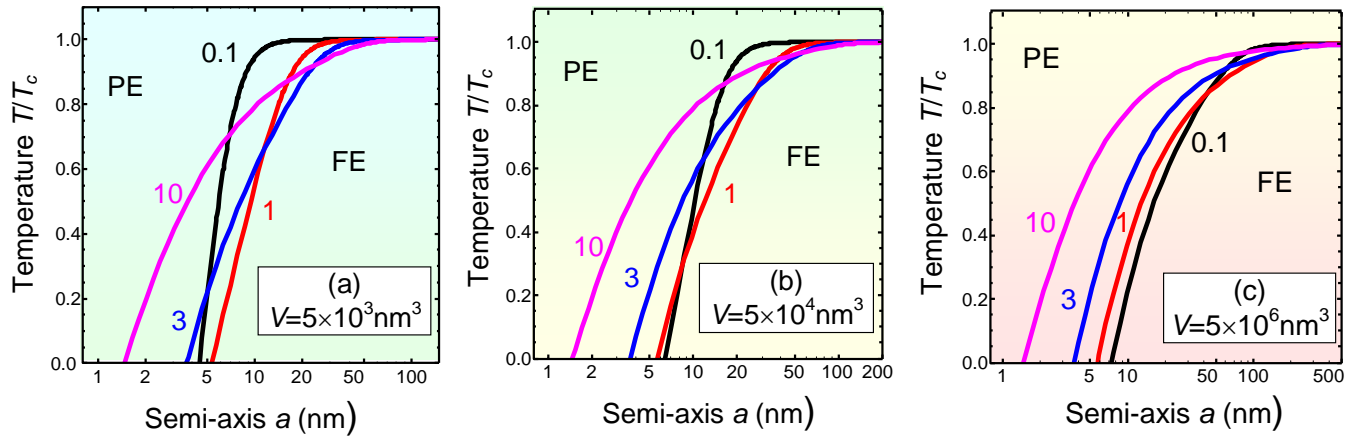

**FIG. 3.** Phase diagrams of BiFeO$_3$ nanoparticles in the coordinates "temperature – length of ellipsoid semi-axis $a$", calculated at the fixed volume $V=5\times10^3$ nm$^3$ **(a),** $V=5\times10^4$ nm$^3$ **(b),** and $V=5\times10^6$ nm$^3$ **(c),** for different aspect ratios $a/b=0.1, 1, 3, 10$ (see numbers near the curves) and the screening length $\lambda=1$ nm.

The transition lines corresponding to different $a/b$ values intersect at all chosen values of $V$ but the intersections are more separated at small volumes [compare **Figs. 3(a), (b)** and **(c)**]. This result is nontrivial and important, since the appearance of such intersections can affect the optimization procedure of the nanoparticles polar properties at a fixed volume, thus being useful for the potential applications of BiFeO$_3$.

Hence it is important to explain the origin of the intersections. Based on the analytical expressions (2)-(6), we conclude that the transition temperatures can be the same for different sizes *a,b,c* of the particle if only the effective depolarization factor $n_d(a,b,c)$ given by Eq.(4) is an ambiguous function of the sizes at fixed volume *V*. Elementary calculations show that $n_d(a,V,\gamma)$ in Eq.(4) is an ambiguous function of *a* and γ at fixed *V*.

The spontaneous polarization dependences on the length of ellipsoid semi-axis *a* calculated for fixed particle volume $V = 5\times10^3$, $5\times10^4$, and $5\times10^6$ nm$^3$ at room temperature are shown in **Figs. 4(a), (b)** and **(c)**, respectively**.** Curves in the each panel are calculated for several values of the aspect ratio *a/b* =0.1, 1, 3, 10 and fixed particle volume *V*. The spontaneous polarization appears at the critical size $a_{cr}(a,V)$ and increases with size *a* for all *a/b* ratios. The critical size decreases with the increase of the *a/b* ratio. The spontaneous polarization saturates to the bulk value ~ 1 C/m$^2$ at sizes *a* > 50 nm. Note that the polarization of the nanoparticles with *a/b* = 10 saturates slower than that of the particles with *a/b* = 0.1. Hence the saturation rate increases with decreasing *a/b*, as can be observed from **Figs. 4(a)-(c)**. The polarization curves have intersection points for *a/b* =0.1 and *a/b* =3 at *a*=5 nm [**Fig. 4 (b)**]; the curves for *a/b* =0.1 and *a/b* =1 also intersect at *a*≈8 nm. Hence the comparison of the curves order shown in **Figs 4** allows us to state that the dependence of the nanoparticle polarization on the semi-ellipsoid *a*-axis length for *V*=5×10$^3$ nm$^3$ is unusual due to the mixed order of different curves and their intersections at small sizes, whereas for *V*=5×10$^6$ nm$^3$ the curves order is expected with increasing *a/b*, and the intersections can appear for long semi-axes only (compare with the diagrams shown in **Figs.3**). The phase transition for the highest aspect ratio *a/b* =10 occurs at the minimum value of *a* (see the magenta curve). The phase transition for the smallest ratio *a/b* = 0.1 (see the black curve) corresponds to the maximal value of *a*.

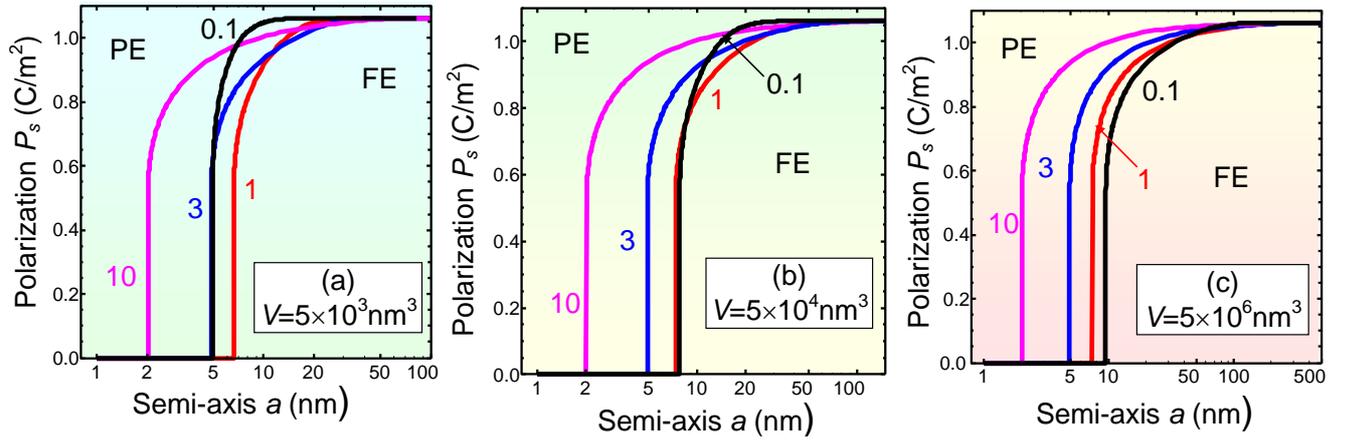

**FIG. 4.** The spontaneous polarization dependence on the length of ellipsoid semi-axis $a$ calculated at room temperature (300 K), fixed volume $V=5\times 10^3$ nm$^3$ **(a)**, $V=5\times 10^4$ nm$^3$ **(b)** and $V=5\times 10^6$ nm$^3$ **(c)** for different aspect ratios $a/b$ =0.1, 1, 3, 10 (see numbers near the curves). Screening length $\lambda$=1 nm, $T$=300 K; other parameters corresponds to BiFeO$_3$ compound.

### IV. PARAMAGNETOELECTRIC (PME) COEFFICIENT AT FIXED VOLUME

Below we consider only the paramagnetoelectric (PME) effect at fixed volume that exists even in a paramagnetic phase and is insensitive to the magnetic symmetry changes in nanoparticles. As it was shown in Refs. [56, 57] the PME coefficient $\eta$ is proportional to the biquadratic magnetoelectric coupling (ME) coefficient $\xi_{MP}$ that couples the second powers of polarization and magnetic order parameters, average spontaneous polarization $P_S(T)$ given by Eq.(3), linear magnetic and dielectric susceptibilities $\chi_M(T)$ and $\chi_{FE}(T)$ [29]:

$$\eta(T) = -P_S(T)\chi_{FE}(T)(\chi_M(T))^2 \xi_{MP} \equiv \begin{cases} \dfrac{-\xi_{MP}(\chi_M(T))^2}{2\sqrt{\alpha_T \beta(T_{cr}(a,b,c)-T)}}, & T < T_{cr}, \\ 0, & T > T_{cr}. \end{cases} \qquad (7)$$

Approximate expression for magnetic susceptibility is taken the same as in Ref. [29], $\chi_M(T) = \dfrac{\mu_0}{\alpha_M^{(T)}(T-\theta) + \xi_{LM}L^2 + \xi_{MP}P_S^2(T)}$. Equation (7) is valid in the FE-AFM phase (with nonzero AFM long-range order parameter $L \neq 0$) as well as in the paramagnetic FE phase without any magnetic order at $L$=0. Parameters $\xi_{LM}$ and $\xi_{MP}$ are the biquadratic ME coefficients, which couple with the second powers of polarization and magnetic order parameters in the ME energy.

The dependences of the PME of the coefficient on the volume of the half ellipsoid V, calculated at room temperature (T = 300 K) for various values of the half-ratio $a/b$ are shown in **Fig. 5**. The values of the other semi-axis c are chosen from the formula $c = 3V/(4\pi ab)$ (see different curves calculated for $a/b$ = 0.1, 2, 3, 10). The PME coefficient is normalized to the bulk value. The PME coefficient is zero for sizes $a < a_{cr}(b,c)$ due to the spontaneous disappearance of polarization, it appears at $a > a_{cr}$ and diverges at a critical value $a = a_{cr}(b,c)$, then decreases with $a$ increase. The PME coefficient reaches a bulk value at sizes $a \gg 100$ nm. The discrepancies at $a = a_{cr}(b,c)$ demonstrate the possibility of obtaining a giant PME effect in BiFeO$_3$ nanoparticles near the size-induced transition from the FE phase to PE phase, in particular, the normalized PME coefficient substantially exceeds 1 for sizes $a_{cr}(b,c) \leq a < 2a_{cr}(b,c)$. Note that the PME coefficient at V = $10^2$ nm$^3$ achieves saturation significantly faster than at V = $10^6$ nm$^3$ [compare the curves in **Figs.5 (a) -5 (c)** ]. PME curves calculated for different V are very close to each other at V = $10^2$ nm$^3$ and are well separated from each other at V = $10^6$ nm$^3$.

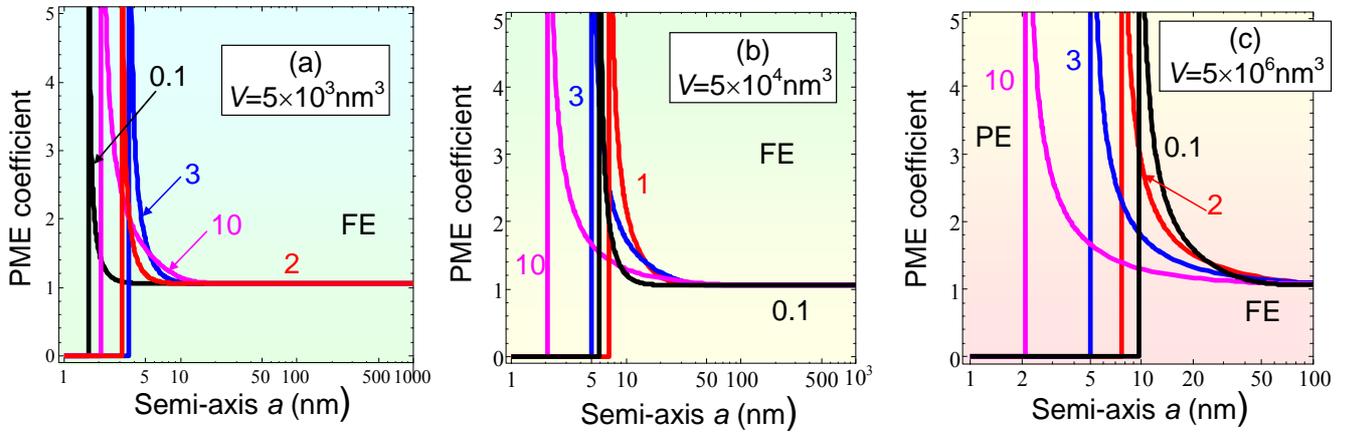

**FIG. 5.** The PME coefficient dependence on the length of ellipsoid semi-axis $a$ calculated at room temperature (300 K), fixed volume $V=5\times10^3$ nm$^3$ **(a),** $V=5\times10^4$ nm$^3$ **(b)** and $V=5\times10^6$ nm$^3$ **(c)** for different aspect ratios $a/b$ =0.1, 2, 3, 10 (see numbers near the curves). Screening length $\lambda$=1 nm, $T$=300 K; other parameters corresponds to BiFeO$_3$ compound.

The comparative analysis of **Fig. 5 (a) - (c)** confirms our conclusion that the size effects for spontaneous polarization and PME coefficient are sensitive to the ratio $bc/a^2$ in the direction of polarization for a given volume and less sensitive to the absolute sizes of the nanoparticles.

### V. CONCLUSION

To resume, using a combination of the Landau-Ginzburg-Devonshire phenomenology, classical electrostatics, and elasticity theory, we have studied the size effect on the phase diagrams and ferroelectric polarization of semi-ellipsoidal BiFeO$_3$ nanoparticles with three different semi-axes and fixed volume $V$. The fixed volume condition corresponds to realistic technological conditions of nanoparticle fabrication [39-41]. Our analysis utilizes the analytical expressions derived in the earlier study [29] for the dependence of the ferroelectric transition temperature, average polarization and PME coefficient on the particle size, and here we account for the fact that the product of the particle's semi-axes $abc$ is fixed, since the semi-ellipsoid volume is $V = (2/3)\pi abc$. The analyses of the obtained results leads to the conclusion that the size effects of the phase diagrams and polarization nontrivially depend on the particle volume $V$ and aspect ratio in the polarization direction $a/b$. These results open the way to control the properties and govern phase diagrams under the realistic experimental conditions of fixed particle volume.

**ACKNOWLEDGMENTS.** This project has received funding from the European Union's Horizon 2020 research and innovation programme under the Marie Skłodowska-Curie grant agreement No 778070. M.V.S. and D.V.K acknowledge MK-1720.2017.8, RFFI (#17-58-45026). A.L.K. acknowledges support from CICECO-Aveiro Institute of Materials (Ref. FCT UID/CTM/50011/2013), financed by national funds through the FCT/MEC and when applicable co-financed by FEDER under the PT2020 Partnership Agreement. A.N.M. work was partially supported by the National Academy of Sciences of Ukraine (projects No. 0117U002612 and No. 0118U003375)